\documentclass{mathincs}

\usepackage{booktabs} 
\usepackage{verbatim}
\usepackage{url}
\usepackage{graphicx}
\usepackage{caption}
\usepackage{subcaption}
\usepackage{csvsimple}
\usepackage{multicol}
\usepackage{textcomp}
\usepackage{svg}
\usepackage[square,numbers]{natbib}
\usepackage{lipsum}
\usepackage{amsfonts}
\usepackage{amsmath}
\usepackage{hyperref}
\usepackage{doi}
\usepackage{cleveref}
\usepackage{todonotes}
\let\oldtodo\todo
\renewcommand{\todo}[1]{\oldtodo[inline]{#1}}

\usepackage{pgfplots}
\usepgfplotslibrary{dateplot}

\newcommand{\RR}{\mathbb{R}}

\DeclareMathOperator*{\operation}{operation}
\DeclareMathOperator*{\aveg}{aveg}

\pgfplotsset{compat=1.18}
 \theoremstyle{definition}
 
 \theoremstyle{remark}

 \numberwithin{equation}{section}

\begin{document}
%
%
%
\title[Datasets and Paradigms in ML for SC]
 {Lessons on Datasets and Paradigms in Machine Learning for Symbolic Computation: A Case Study on CAD}

\author[Tereso {del R{\'i}o}]{Tereso {del R{\'i}o}}
\address{Coventry University, UK}
\email{delriot@coventry.ac.uk}

\author[Matthew England]{Matthew England}
\address{Coventry University, UK}
\email{Matthew.England@coventry.ac.uk}



\date{January 19, 2024}

\begin{abstract}
Symbolic Computation algorithms and their implementation in computer algebra systems often contain choices which do not affect the correctness of the output but can significantly impact the resources required:  such choices can benefit from having them made separately for each problem via a machine learning model.  
This study reports lessons on such use of machine learning in symbolic computation, in particular on the importance of analysing datasets prior to machine learning and on the different machine learning paradigms that may be utilised.  We present results for a particular case study, the selection of variable ordering for cylindrical algebraic decomposition, but expect that the lessons learned are applicable to other decisions in symbolic computation.  

We utilise an existing dataset of examples derived from applications which was found to be imbalanced with respect to the variable ordering decision.  We introduce an augmentation technique for polynomial systems problems that allows us to balance and further augment the dataset, improving the machine learning results by 28\% and 38\% on average, respectively. We then demonstrate how the existing machine learning methodology used for the problem $-$ classification $-$ might be recast into the regression paradigm.  While this does not have a radical change on the performance, it does widen the scope in which the methodology can be applied to make choices.
\end{abstract}

\maketitle

\maketitle

\section{Introduction}
\label{sec:Intro}

Symbolic computation algorithms, including those commonly used within theory solvers for SMT, often have within them a variety of choices to be made: choices that do not affect the correctness of the outputs, but can still have a significant effect upon the resources needed for such algorithms to return an output and on the form that such outputs take.  For example, when running Buchberger's algorithm, we may choose any $S$-pair of polynomials from the list of those to process when computing the next S-polynomial: as long as we process all pairs in the list eventually, then we obtain a Gr\"{o}bner basis of the original polynomials. However, it is well observed that the strategy utilised to make such choices can significantly impact the cost of running the algorithm, and also how the final output is presented.  Historically, these choices have been made using heuristics developed by experts, such as the sugar heuristic developed to choose pairs of polynomials in Buchberger's algorithm \cite{gioviniOneSugarCube1991}.  However, not all choices are well-documented or evaluated openly in the literature.

There has been a trend in recent years for programmers to train machine learning models to make these choices. \emph{Machine Learning} (ML) refers to statistical techniques that learn rules from data.  It has been well documented that ML has outperformed expert humans in a wide range of fields, driven forward by growth in computing power, new techniques, and methods, and an explosion in available training data. It was first suggested a decade ago that ML might enhance the effectiveness of symbolic computation algorithms by replacing human-designed heuristics \cite{huangApplyingMachineLearning2014}. This could allow non-expert users of symbolic computation to optimise algorithms to their application domain, and for mathematicians to focus on the theory with less distraction from such implementation details.  Further, the emergence of Explainable AI techniques may allow for ML to provide suggestions and guidance on algorithm development, even for those who do not wish to incorporate ML into their final software \cite{PdREC24}.

We will investigate two aspects of ML methodology, as introduced in the next two subsections, in the context of choices for symbolic computation.  We will present experimental results for our case study problem of choosing the variable ordering for cylindrical algebraic decomposition, introduced shortly, but emphasise that the lessons drawn should be applied more widely. 

\subsection{Machine learning paradigms}
\label{subsec:intro_paradigms}

Many of these choices in symbolic computation fall naturally into the ML \emph{classification} paradigm: where we train the ML model to choose one option from a set of discrete possibilities.  An alternative paradigm is ML \emph{regression}:  where we train the ML model to estimate a continuous real numbered variable.  These are often presented as distinct approaches for tackling different types of problem, but this paper notes that either may be used for optimisation of symbolic computation algorithms.  Given our heuristic choice from a set of distinct possibilities, we identify the optimal one by the evaluation of some metric, e.g. the time taken for the algorithm to run with that choice, or the size of the output produced with that choice.  This metric is used to identify the best choice in the classification problem; but it could also be the variable we seek to estimate in the regression problem, with the set of estimations for the different possibilities then used subsequently to make that discrete choice.  

This raises the question of which paradigm is better to view such problems.  Regression is certainly a more difficult problem than classification; however, for our purposes, the learning needs only to be sufficient to \emph{rank} the discrete possibilities allowing scope for significant error in the estimations of the continuous variable without loss of performance in the symbolic computation.  With that in mind, our hypothesis is that regression should be superior since it exposes more information to the ML model during training.  Given a choice between $n$ discrete possibilities, the classification paradigm treats each example as a single instance for the model to learn from, while the regression paradigm treats each example as $n$ different instances for the model to learn from.  The regression paradigm should thus allow the ML model to obverse more information: not just which choice was the best, but what effect all of the possible choices have.  

\subsection{Data augmentation}
\label{subsec:intro_data_augmentation}

Data augmentation consists of generating new data instances from existing ones. It is a widely-used technique in ML applications where a larger dataset can help tackle overfitting and increase the accuracy of the resulting model. Moreover, it can be used to mitigate biases in the dataset and to reduce the cost of labelling \cite{shortenSurveyImageData2019}.  

Data augmentation is commonly used to generate new images in Computer Vision ML applications in particular.  For example, in a vision classification problem we can rotate and reflect the image without changing its label. In our work later, we will demonstrate how these ideas of data augmentation may be used in the context of choices for symbolic computation through the permutation of the variables.  In our case, the label of these new instances will change but can be easily identified from the original label: important since the labelling of data with symbolic computation will often be the most expensive part of training a ML model to make such choices.

\subsection{Our case study problem: variable ordering for CAD}
\label{subsec:intro_CAD}

This paper will focus on the case study of choosing the variable ordering for Cylindrical Algebraic Decomposition (CAD): an algorithm that decomposes the $n$-dimensional real space into distinct cells (connected regions of the space).  It is traditionally applied on sets of polynomials to produce cells in which each of those polynomials is sign-invariant (either positive, negative, or zero throughout the cell). Such a CAD is a powerful tool for analysing and understanding the behaviour of polynomial systems, enabling us to solve tasks such as real quantifier elimination. However, the doubly exponential complexity of CAD with respect to the number of variables strongly limits its practical application.  The careful study of its optimisation can only add new applications to its scope.

One particular area in need of optimisation is the variable order.  This may be free or constrained, depending on the intended use of the CAD after it is created.  When free, it has been observed that the choice of variable ordering can significantly impact both the theoretical complexity of CAD \cite{brownComplexityQuantifierElimination2007} and the practical tractability of CAD implementations \cite{dolzmannEfficientProjectionOrders2004}.  
Much of the experimental methodology used in this paper is replicated from our prior work on this topic, with the changes under study being the paradigm used to train the ML models and the augmenting of the dataset. 

\subsection{Paper outline}
\label{subsec:intro_outline}

\Cref{sec:LR} contains a literature review: we start with previous work on the use of ML in symbolic computation, then describe the CAD algorithm used as a case study in this paper and the current state-of-the-art approaches to choose its variable ordering. \Cref{sec:methodology machine learning} then covers some common methodology used in all our experiments.  

\Cref{sec:Classification} considers the classification paradigm, replicating previous work and identifying weaknesses that arise from the imbalance of the dataset. 
We describe how data augmentation fixes both these issues and improves the performance of ML models. 
This contribution was originally presented in a paper of the SC-Square 2023 workshop \cite{delrioDataAugmentationMathematical2023} which this present article extends. 

In \Cref{sec:RegressionOrderings} we move on to reframe this problem in the regression paradigm, comparing to classification, and in \Cref{sec:regression variables} we present an alternative formulation that is more widely applicable.   Finally, in \Cref{sec:conclusion} we share our conclusions obtained with these experiments.

\section{Literature Review}
\label{sec:LR}

ML has been improving multiple fields over the past decades. In some safety-critical areas, such as automatic driving and medicine, ML models are not fully trusted, as they are not 100\% accurate and an error could be fatal.  However, they are still used to assist humans, such as when your car applies lane assist technology, or when a doctor's attention is drawn to particular cases.  Similarly in mathematics, and thus symbolic computation, we usually require exact answers and so on the surface it may seem unlikely to use ML in computer algebra systems; except perhaps for drawing human attention to things of potential interest, as in \cite{DVBBZTTBBJLWHK21}, or where an ML-produced output can be easily checked for correctness, as in \cite{lampleDeepLearningSymbolic2020}.  
However, as discussed in the introduction, symbolic computation algorithms often contain choices that do not affect correctness but which have a significant impact on the resources needed: such choices may be safely addressed by ML and are the focus of our study.  We start by reviewing the prior literature for examples of this, paying particular attention to the learning paradigm utilised, before reviewing the literature for our case study problem. 

\subsection{Machine learning in symbolic computation}
\label{sec:literature review machine learning}

\subsubsection{Traditional classification paradigm}
\label{subs:LRTrad}

Most of the examples in the literature of ML to optimise symbolic computation take the form of supervised learning with the classification paradigm: we will survey those here.  \emph{Supervised} ML means that we start with a dataset of labelled instances that we use to train the ML models.  For example, a dataset of images where each image is labelled as either a cat or a dog is used to train a binary classifier to make that distinction. It is important that the data structure has some fixed features across the instances in the dataset, e.g. images of the same resolution or a representation as a fixed feature vector.  

This paradigm was first used for symbolic computation by Huang et al.~in 2014 \cite{huangApplyingMachineLearning2014}.  They trained a support vector machine to choose the variable ordering for three-variable CAD problems (meaning six possible variable orderings); and later the same authors used this methodology to decide when to precondition a CAD input with computation of a Groebner basis \cite{HEDP16}, a binary classification. Florescu and England built on this, experimenting with models \cite{englandComparingMachineLearning2019} and features \cite{florescuAlgorithmicallyGeneratingNew2019}. 

There are further examples of supervised ML classification for computer algebra away from CAD in the literature.  An early example was by Simpson et al. who used ML classifiers to pick from different algorithms to compute the resultant \cite{simpsonAutomaticAlgorithmSelection2016}, substantially improving the runtimes in both Maple and Mathematica.  Recently, Barket et al. have experimented with ML to select the order in which to try 12 possible sub-algorithms for symbolic integration in Maple \cite{BEG24}. They experimented with both traditional LSTMs (Long-Short Term Memory neural networks) which process sequential data and a variant, Tree LSTMs, which processed the input as expression trees.  They found the latter structure beneficial, and outperformed the standard Maple mechanism to select the sub-algorithm.

ML classifiers to select the sub-algorithm for a task have also proved powerful in computational logic: where SATZilla was an early example to select a SAT-solver for a problem \cite{xuSATzillaPortfoliobasedAlgorithm2008}, and recently MachSMT did this for the far more diverse world of SMT \cite{scottMachSMTMachineLearningbased2021}.

Other recent examples of ML classification in computer algebra include the use of classifiers to predict features of amoebae in algebraic geometry \cite{BHH23}, and the number of real solutions to polynomial systems \cite{BHMRT23}.  However, we draw a distinction between these and the prior examples since here ML is directly predicting the output, and thus inaccuracy in the ML predictions does not just lead to inefficiencies, but to incorrect answers. 

\subsubsection{Modified classification paradigms}

Brown and Daves used ML to choose a polynomial ordering for the NuCAD algorithm in \cite{brownApplyingMachineLearning2020}. This required them to be able to choose between an \emph{a priori} unspecified number of options: i.e., it is not just that there are multiple options but that the number of options may differ at each selection point.  Brown tackled this by training a binary classifier to choose between two polynomials and then having the options compete with each other under this classifier until one stands as a winner.

The work in the literature that comes closest to our ideas in \Cref{sec:RegressionOrderings} is that of Florescu and England \cite{florescuImprovedCrossValidationClassifiers2020}. They also started from the recognition that the classification paradigm was not revealing as much information to the ML models as there was available.  They edited the procedure that performs cross-validation for hyperparameter selection on the classifiers so that the hyperparameters were selected based, not on classification accuracy of choices, but on the runtime achieved by those choices.  In fact, this improved performance, motivated our hypothesis.  However, after the selection of hyperparameters was completed, the actual classifiers in \cite{florescuImprovedCrossValidationClassifiers2020} were still trained on accuracy as normal and so this does not represent a full use of the regression paradigm.

\subsubsection{Regression and reinforcement learning}

There are fewer examples of regression for the optimisation of computer algebra in the literature.   
We were particularly motivated by the work of Peifer et al.~which used regression to choose $S$-pairs of polynomials in Buchberger's algorithm \cite{peiferLearningSelectionStrategies2020}. Classification could not be used here since the number of possible $S$-pairs to choose from differed from one problem instance to another.  They instead trained a regressor to predict the number of polynomial additions that are required after choosing a given pair, in an iteration of the main loop of Buchberger's algorithm. In this way, instead of informing the algorithm about the best option, the algorithm is informed about how much work the chosen option required.  

As well as a move to regression instead of classification, this work also represents a shift from supervised to unsupervised learning.  Peifer et al.~did not measure the number of additions required for every possible $S$-pair in advance to create a dataset to learn from.  Instead, their model made choices, got scored (the score being the number of additions), and amended its parameters in each iteration of the algorithm.  This approach is usually known as \emph{reinforcement learning}\footnote{Another example of reinforcement learning is \cite{KM20} which used it to learn the best pivot to use in Gaussian Elimination.}.   

We note that \cite{peiferLearningSelectionStrategies2020} aimed to minimise the cost of each step, in the hope that if the cheapest decision is taken at each step, then the cost of the overall algorithm will be small.  
There is a risk with this approach that it leads to more $S$-pairs being generated overall, but \cite{peiferLearningSelectionStrategies2020} reported strong performance, with their reinforcement learning agent outperforming all human-designed $S$-pair selection strategies.  This motivates our own hypothesis that regression can allow for deeper learning than classification.  However, unlike \cite{peiferLearningSelectionStrategies2020}, we will not switch to reinforcement learning.  We will directly compare supervised learning with classification and supervised learning with regression to see what, if any, performance difference there is.  

Lastly, as the paper was being finalised, \cite{jiaSuggestingVariableOrder2023} was published with the first use of reinforcement learning to choose the variable ordering for CAD.  Their methodology applied this approach with graph neural networks (using the same graph embedding discussed later in Section \ref{subsubsec:other}).

\subsection{Cylindrical algebraic decomposition}
\label{sec:CAD}

\subsubsection{Definition}

Cylindrical Algebraic Decomposition (CAD) is both a mathematical object and an algorithm to produce such objects, first proposed by Collins in 1975 \cite{collinsQuantifierEliminationReal1975}. It takes a set of polynomials $S_n \in \mathbb{R}[x_1, \dots, x_n]$ as input, along with a variable ordering, and then builds a decomposition of $\mathbb{R}^n$ into connected regions called cells that are each sign-invariant for each of the input polynomials. Further, each cell is semi-algebraic, meaning that it may be described using polynomial constraints; and the cells are arranged in cylinders, meaning the projection of any two cells onto a lower-dimensional space (with respect to the variable ordering) is either equal or distinct.  The latter condition makes membership testing, projection, and negation easy which all aid the use of CAD for quantifier elimination; the former condition allows for easy solution formula creation.

\subsubsection{Algorithm}

CAD algorithms usually proceed in two phases: projection and lifting. The first step of projection takes the set of polynomials $S_n$ and returns a set of polynomials $S_{n-1}$ without the largest variable in the ordering, $x_n$. In the next step of the projection phase, a set of polynomials without $x_{n-1}$ is created. This phase continues until we reach a set of univariate polynomials in $x_1$.

In the first step of the lifting phase, a CAD sign-invariant for the set of polynomials $S_1$ is built.  This may be done by using real root isolation and forming cells from the roots and intervals inbetween. In the next step, this first CAD of $\mathbb{R}$ is extended to a CAD of $\mathbb{R}^2$ sign-invariant for the set of polynomials $S_2$. This is achieved by taking each cell $C \in \mathbb{R}$ and choosing a single sample point $s \in C$ that we substitute into $S_2$ to produce univariate polynomials upon which we can apply real root isolation.  We are able to conclude the sign-invariant decomposition we form for $\mathbb{R} \times s$ as representative for $\mathbb{R} \times C$, meaning the same split into cells according to roots of polynomials and the intervals inbetween throughout.  The conclusion may be drawn because the information captured in the projection phase encodes the places where the behaviour of those roots changes (the notion of \emph{delineability}).  This lifting phase continues until a CAD sign-invariant decomposition of $\mathbb{R}^n$ is built.

\subsubsection{Complexity}

Davenport and Heinz proved that CAD generates a doubly exponential number of cells ($2^{2^n}$) in the worst case with respect to the number of variables ($n$) \cite{davenportRealQuantifierElimination1988}. 
Brown and Davenport in \cite{brownImprovedProjectionCylindrical2001} have shown that there is a family of sets of polynomials for which the right choice of variable ordering would yield a constant complexity and the wrong choice of variable ordering would result in a doubly exponential complexity. Numerous practical experiments have also shown the significant effect of the variable ordering choice in practice, e.g. \cite{dolzmannEfficientProjectionOrders2004, huangApplyingMachineLearning2014}.  This is the choice that we seek to optimise.

\subsection{Heuristics for choosing the CAD variable ordering}
\label{sec:heuristics}

We use \emph{heuristic} to refer to a rule we may apply to choose a CAD variable ordering.  Heuristics are not guaranteed to produce an optimal answer, but they should produce an answer quickly.  They are usually based on a simple feature (metric) or combination of features of the input.  Although commonly designed by humans, we may also think of an ML model as a heuristic (noting that while ML models may be expensive to train, they are, once trained, quick to make a prediction on a single instance).  In this section, we will present the best non-ML heuristics that have been found for our problem so far.

\subsubsection{The Brown heuristic}

For many years, a well-known and widely used heuristic is the one of Brown introduced in his ISSAC 2004 tutorial notes \cite{brownCompanionTutorialCylindrical2004}. Denoted \texttt{Brown}, this uses three simple criteria in turn, breaking ties in previous ones with the subsequent ones.  For a given set of polynomials $S$, it projects first:
\begin{itemize}
      \item $x_i$ s.t. $\max\limits_{p\in S}\left(\max\limits_{monom \in p}\left(degree_{x_i}\left(monom\right)\right)\right)$ is minimal,  
      \\(i.e. using the highest degree with which the variable appears in the polynomials); \\ breaking ties with,
      \item $x_i$ s.t. $\max\limits_{p\in S}\left(\max\limits_{monom \in p}\left(totaldegree_{x_i}(monom)\right)\right)$ is minimal, 
      \\ (i.e. using the highest total degree of a monomial in which the variable appears); \\ breaking ties with
      \item $x_i$ s.t. $\max\limits_{p\in S}\left(\max\limits_{monom \in p}\left(sign(degree_{x_i}(monom))\right)\right)$ is minimal,  
      \\ (i.e. using the number of monomials in which the variable appears).
\end{itemize}
It is not specified in \citep{brownCompanionTutorialCylindrical2004} what to do if there is a third tie: our implementation randomly picks between the tied variables. Furthermore, whether the entire ordering of variables is chosen at once or one variable at a time taking into account intermediate projections is also not specified in \citep{brownCompanionTutorialCylindrical2004}. 

\subsubsection{Other human-designed heuristics}
\label{subsubsec:other}

There have been other human-designed heuristics since then that used more expensive algebraic information (see \cite{dolzmannEfficientProjectionOrders2004}, \cite{BDEW13}, and \cite{wilsonUsingDistributionCells2015}).  However, given our emphasis on heuristics that give answers quickly and the fact that they do not produce results much better than \texttt{Brown} \citep{HEWBDP19}, we do not consider them further here.

More recently, the authors of this article proposed the heuristic \texttt{gmods} in \cite{delrioNewHeuristicChoose2022}, motivated by the complexity analysis of CAD. This heuristic projects first the variable with the lowest degree in the product of the polynomials.   Again, if there are ties, our implementation picks randomly.

\begin{figure}[ht]
    \centering
    \includegraphics[width=0.3\textwidth]{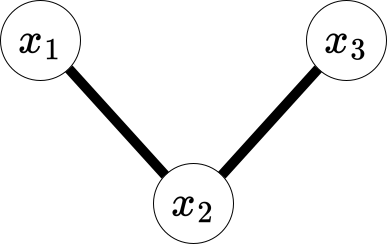}
    \caption{Graph associated to $S_G$.}
    \label{fig:chordal example}
\end{figure}

Another recent heuristic was proposed by Li et al. (2021) \cite{liChoosingVariableOrdering2021}.  This involved a graph structure associated with a polynomial set to capture sparsity information:  the nodes of the graph are the variables in the polynomial set, with two nodes connected if they appear in the same polynomial. For example, the set of polynomials $S_G=\{x_1^3x_2-x_1+2, x_2^4-x_3\}$ would be associated with the graph in \Cref{fig:chordal example}.  Their heuristic uses operations upon this graph to identify a variable ordering that seeks to preserve variable sparsity.

\subsubsection{Human-level heuristics}

The state-of-the-art human-level heuristic in the current literature is \texttt{T1}, from \cite{PdREC24}.  We call this \emph{human-level} because although it was informed by ML techniques, it can be expressed independently of an ML model in a similar quantity of text as the human-designed heuristics above.  In fact, it uses three simple features in a chain just like \texttt{Brown}.  The features were those found most impactful after a SHAP analysis to interpret decisions of ML classifiers trained to choose the best variable ordering \cite{PdREC24}. For a given set of polynomials $S$, this heuristic projects first:
\begin{itemize}
      \item $x_i$ s.t. $\sum\limits_{p\in S}\left(\max\limits_{monom \in p}\left(degree_{x_i}\left(monom\right)\right)\right)$ is minimal,  
      \\(i.e. using the degree of the variable in the product of the polynomials); \\ breaking ties with,
      \item $x_i$ s.t. $\aveg\limits_{p\in S}\left(\aveg\limits_{monom \in p}\left(degree_{x_i}(monom)\right)\right)$ is minimal, 
      \\ (i.e. using the average of the average degree of the variable in the polynomials); \\ breaking ties with
      \item $x_i$ s.t. $\sum\limits_{p\in S}\left(\sum\limits_{monom \in p}\left(degree_{x_i}(monom))\right)\right)$ is minimal,  
      \\ (i.e. using the sum of all degrees of the variable in the polynomials).
\end{itemize}
Once again, if there is a third tie our implementation picks randomly.

\section{Our Methodology for Machine Learning}\label{sec:methodology machine learning}

In this section, we present the core of the ML methodology we will follow in all the experiments of the following sections.  

\subsection{Dataset}\label{sec:dataset}

To train and test supervised ML models, one needs a dataset of labelled instances to train upon. For our purpose, each instance is a set of polynomials in three variables, labelled with the best variable ordering for CAD (the one for which a CAD could be built in the least time). We build this as follows.

\begin{itemize}

\item \textbf{Acquisition of sets of polynomials:} We select the 5942 three-variable problems from the \texttt{QF\_NRA} category of the SMT-LIB \cite{barrettSatisfiabilityModuloTheories2018}, and extract the set of polynomials used from each.  We acknowledge that these are all satisfiability problems and so do not represent the full application range of CAD (e.g. the more general problem of quantifier elimination).  However, they do (mostly) arise from applications, making performance upon them particularly meaningful.  Most problems are generated by the theorem prover MetiTarski \citep{paulsonMetiTarskiFuture2012}, but there are also examples coming from, e.g., biology \citep{bradfordIdentifyingParametricOccurrence2020} and the GeoGebra dynamic geometry tool \cite{vajdaGeoGebraRealgeomReasoning2020}.

\item \textbf{CAD timings:} For each set of polynomials, \texttt{CylindricalAlgebraicDecomposition} from Maple's Regular Chains Library \citep{CM14a} is timed to build the CAD for all six possible variable orderings. A time limit of 30 seconds is used and for the cases in which no variable ordering finished within the limit, the process was repeated with a time limit of 60 seconds.

\item \textbf{Removal of over-complicated problems:} Problems in which the construction of a CAD took more than 60 seconds for all six orderings are discarded, leaving 5599 problems.

\item \textbf{Elimination of duplicates:} Each problem has associated with it six CADs (one for each ordering) and thus a vector of six CAD cell counts.  When we identify two problems for whom these vectors are identical then we discard one problem.  This happens extensively in the dataset where there are many problems that differ only by a single coefficient, having little effect on the geometry/topology of the problem.  After this step, only 1019 problems are left.  We note that this step is the main difference between our work and \cite{hesterAugmentedMetiTarskiDataset2023}:  despite the latter using six times more data, they reported similar performance of the ML models trained.  Aside from efficiency, this should avoid data leakage between testing and training.
\end{itemize}

\subsection{Data embedding via features}
\label{sec:creating embeddings}

There are various possible embeddings that ML models can take as input (e.g. images, text, numbers), but there is no ML model known to the authors that takes a set of polynomials directly as an input.  In line with prior work on ML for the CAD variable ordering problem, we take a feature-engineering approach, representing each set of polynomials by a vector of floats defined from simple-to-compute metrics on the input such as those used by \texttt{Brown} and \texttt{T1} above \cite{huangApplyingMachineLearning2014, florescuAlgorithmicallyGeneratingNew2019, chenVariableOrderingSelection2020}.

Florescu and England \cite{florescuAlgorithmicallyGeneratingNew2019} proposed a semi-automated approach which allowed them to generate 1728 features for each variable in a given set of polynomials in three variables. However, many of these features were copies of each other, and there are 27 per variable which are unique.  
We will imitate their approach using different, hopefully more accessible, notation.  Let us first introduce four different ways to convey information about a polynomial with respect to a variable. We will use the variable $x_1$ in the polynomial $f=2x_1^3x_2+x_1^2x_2x_3+2x_1^2x_3^3-3x_1-x_2^3x_3-4x_3^2+7$ as an example.
\begin{enumerate}
    \item Extract the degree of the variable of each monomial. E.g. $I_{1,x_1}(f)=[3,2,2,1,0,0,0]$.\footnote{The order of monomials in a polynomial is not really relevant but is captured by the list data-structure:  however, this order does not affect any of the operations performed upon the list later.}
    \item Extract the total degree of each monomial in which the variable appears. \\ E.g. $I_{2,x_1}(f)=[4,4,5,1]$.
    \item Represent with a \textquotesingle$1$\textquotesingle\ for each monomial containing the variable and with a \textquotesingle$0$\textquotesingle\ from each monomial not containing the variable. E.g. $I_{3,x_1}(f)=[1,1,1,1,0,0,0]$.
    \item Extract the total degree of each monomial in which the variable appears and add a zero for each monomial in which the variable does not appear. E.g. $I_{4,x_1}(f)=[4,4,5,1,0,0,0]$.
\end{enumerate}
Information relative to a variable in a set of polynomials is then described with a vector of vectors. For example, using the first option above, the information relative to the variable $x_1$ in the set of polynomials $S=\{4x_1^3x_3-x_2x_3+5x_3^2-1, f\}$ is $I_{1,x_1}(S)=[[3,0,0,0],[3,2,2,1,0,0,0]]$.

To condense this vector of vectors down to a single feature, Florescu and England \cite{florescuAlgorithmicallyGeneratingNew2019} proposed using the operations of maximum, summation, and average, taking inspiration again from the features in the human-design heuristic of Brown \cite{brownCompanionTutorialCylindrical2004}. Taking the example in the previous paragraph, an option would be using summation and then average, to obtain $$avg(sum(I_{1,x_1}(S)))=avg([sum([3,0,0,0]), sum([3,2,2,1,0,0,0])])=avg([3,8])=5.5.$$

The template that Florescu and England \cite{florescuAlgorithmicallyGeneratingNew2019} proposed to extract features relative to the variable $x_i$ with respect to the set of polynomials $S$ can be simplified to
\[
\operation\limits_{p\in S}\left(\operation\limits_{M \in p}\left(I_{j,x_i}(S)\right)\right)
\]
where $\operation$ is either a maximum, a sum, or an average, and where $j\in \{1,3,4\}$.

This allowed them to extract nine features per variable for each $j$. We note that  \cite{florescuAlgorithmicallyGeneratingNew2019} used $I_{1,x_i}(S), I_{3,x_i}(S)$ and $I_{4,x_i}(S)$ while we used $I_{1,x_i}(S), I_{2,x_i}(S)$ and $I_{3,x_i}(S)$ (which we found led to better performance).  
This obtains 27 features for each variable $x_i$.  The number is reduced to 26 after $max(max(I_{3,x_i}(S)))$ is eliminated (which we do since it was evaluated to $1$ for every instance in our dataset: every variable $x_i$ appears in every set of polynomials $S$ at least once).

Finally, recall the work in \cite{liChoosingVariableOrdering2021} discussed in Section \ref{subsubsec:other} earlier which utilised a graph embedding of the polynomial set.    
Inspired by their work, we add the degree of the variable in such a graph as a new feature, bringing the total back to 27 features per variable. In the example from Section \ref{subsubsec:other} this new feature would evaluate to 1 for $x_1$ and $x_3$, and to 2 for $x_2$.

\subsection{Dataset preprocessing}
\label{sec:preprocessing}

To prepare the data for the ML models, it is common to standardise it by setting the mean of each feature to 0 and the standard deviation to 1. Standardisation of the data helps avoid biases towards certain features and improves the accuracy and effectiveness of the model, as explained by \cite{kuhnAppliedPredictiveModeling2013}.  
We use the function \texttt{ StandardScaler} from the library \texttt{sklearn.preprocessing} for this.

\subsection{Train/test split}\label{sec:train test split}

We divide our dataset into five separate parts, known as ``\emph{folds}''. We are careful to ensure that problems from the same source (the same directory of the SMT-LIB) end up in the same fold. This is to prevent ``\emph{data leakage}'': a situation in which a model is tested using the same or similar data to the data it was trained on.  
We choose the hyperparameters and train models on four of these folds using the strategy being studied; then the models will be tested on the remaining fold. Repeating this until all folds have been used for testing, and taking the average to get final results.  
This process ensures that our testing is fair and that a model does not benefit from the test/train split, while still ensuring that the models are evaluated on data they have not seen during the selection of its hyperparameters or its training.

\subsection{Hyperparameter tuning}\label{sec:hyperparameter tuning}

In ML, ``\emph{hyperparameters}'' are settings that govern the architecture and learning process of the model. Hyperparameters are set prior to training, in contrast to model parameters which are optimised during training.  Examples of hyperparameters would include the number of layers in a neural network or the number of branches in a decision tree.  

To find the best hyperparameters, we employ a technique called ``\emph{Bayesian Optimisation}'' through the function \texttt{BayesSearchCV} in \texttt{scikit-optimize}. This function is more sophisticated than the more commonly used \texttt{GridSearchCV} which would evaluate all combinations of values for hyperparameters taken from discrete sets, and \texttt{RandomisedSearchCV} which randomly samples from such a grid.  \texttt{BayesSearchCV} tries a fixed number of parameter samples from specified distributions 
and intelligently navigates the hyperparameter space to select these.

Additionally, instead of relying on default metrics like accuracy or mean squared error, we utilise a custom scoring function. Our custom scorer evaluates a model based on the time it takes to construct CADs using the chosen variable ordering, rather than, for example, how often the optimal variable ordering is chosen. This aligns the evaluation used to choose the hyperparameters with the specific goals of our research, similarly to \cite{florescuImprovedCrossValidationClassifiers2020}.

\subsection{ML models used}\label{sec:training models}

During our experiments, we train a variety of models to ensure that the conclusions drawn generalise. The models we use are: 
K-Nearest Neighbours (KNN), 
Random Forest (RF), 
Support Vector Machine (SVM), 
Multi-Layer Perceptron (MLP), and 
Extreme Gradient Boosting (XGB).  For the latter we use the implementation in the \texttt{xgboost} library and for the others the implementations in the \texttt{sklearn} library.  
Models are trained using the \texttt{.fit()} method in the \texttt{sklearn} library with default settings, maintaining a standardised process to fairly compare the changes under scrutiny. 

\subsection{Testing the models}\label{sec:testing models}

Unless otherwise specified, the dataset used to test the different strategies is the \emph{balanced dataset} described later in \Cref{sec:improving datasets} which best ensures the fairness of the comparisons.

In this paper, we aim to find a methodology to make good choices in symbolic computation. In particular, we focus on the case study of choosing good variable orderings.
But what does ``\emph{good}'' stand for in this context? 
To measure the complexity of a CAD we use the time needed to generate it, but as it is impossible to obtain a model that guarantees choosing the optimal ordering every single time we must find ways to compare strategies that may be better in some instances but worse in others.  
We make use of the following metrics.

\subsubsection{Number of solved instances}\label{subsec:solved instances metric}

The most basic metric is the number of solved instances, i.e. the number of instances for which the strategy chose a variable ordering that did not timeout. 

\subsubsection{Time accuracy}\label{subsec:time accuracy}

Time accuracy is the percentage of instances for which the strategy selects the fastest variable ordering to build a CAD.
Accuracy is very commonly used to measure the performance of ML models in classification problems. However, it does not suit our interests very well. 
Imagine that the best variable ordering to build a CAD takes 12.43 seconds. If our strategy chooses a variable ordering that takes 12.47 seconds, that is still pretty good, but the accuracy measure does not see it that way. It treats this almost-right choice as completely wrong, the same way it would treat a terrible choice, even if that choice timed out.  This is even worse when we consider that slight differences in the timings may be the result of computational noise.

\subsubsection{Total time}\label{subsec:total time}

Perhaps the metric most aligned with user satisfaction is total time: the time needed to build a CAD for each instance in the testing dataset, when choosing the ordering according to the strategy being studied.  
Unlike accuracy, this metric treats suboptimal choices based on how suboptimal they are. In the example of \Cref{subsec:time accuracy}, a choice that takes 12.43 seconds will be evaluated only marginally better than a choice that takes 12.47 seconds.

However, this metric does not fairly reflect performance in instances where building a CAD took a small amount of time for all orderings. For instance, if two variable orderings take 1 second and 4 seconds, respectively, to build a CAD, the impact of this choice is equivalent to the impact of a choice between a variable ordering that took 31 seconds and one that took 34 seconds. However, you may argue that selecting an option that costs 4 instead of 1 is a much worse decision than choosing an option that costs 34 instead of one that costs 31.

Moreover, this metric is very dataset-dependent, meaning that a total time of 1000 seconds might be a fabulous time for one dataset while being a terrible time for another dataset.  Therefore, it is a poor metric for evaluating a single sample or a dataset for which the performance of other strategies is unknown. This implies that it is not the right metric to use in, say, Reinforcement Learning, where the model should be rewarded or penalised based on its performance in a group of instances.

\subsubsection{Time markup}\label{subsec:time markup}

We desire a metric that does not completely penalise suboptimal orderings while reflecting the effectiveness in fast instances.
For this purpose, in \cite{delrioNewHeuristicChoose2022}, we proposed \emph{time markup}. This metric is the percentage of extra time needed to build a CAD using the chosen variable ordering measured against the optimal ordering.
Given a chosen suboptimal ordering with time $t_{chosen}$  in a sample where the optimal ordering time is $t_{optimal}$ the markup is defined as 
$$\frac{t_{chosen}-t_{optimal}}{t_{optimal}+1}.$$
We add +1 to the denominator to avoid high markups for small differences in simple problems because these may be caused by computational noise rather than by the orderings themselves.

Over a dataset we define the markup as the average of the markups for all instances.

\subsubsection{Metrics based on cell counts}

An alternative set of metrics to those in Sections \ref{subsec:time accuracy} $-$ \ref{subsec:time markup} could be defined using a count of the number of cells of a CAD rather than the time taken to compute the CAD.  In theory, such metrics would be implementation independent (although in practice, all CAD implementations differ on some of the underlying theory they implement).  The cell count and the time taken should be lightly correlated (at least before any post-processing of cells): we choose to focus on the time related metrics, as it is the measure of most value to users.

\subsubsection{Timeout penalisation}

Note that total time and time markup metrics depend on knowing the time needed to complete a CAD using the variable ordering suggested by the strategy studied. However, as described in \Cref{sec:dataset}, CADs in Maple were called with a time limit, after which, if the call has not finished, it is stopped by Python\footnote{We make each CAD call in a separate Maple session launched and timed by Python to avoid any instance having the benefit of Maple's previous caching of intermediate results.   For example, the same resultant may be needed for two different variable orderings, and we want to ensure the cost of its computation is included in both instances.}. 
When this happens, we do not get to know the time needed to build a CAD with such an ordering; we only know it is more than the chosen time limit. However, we need to choose a numerical value to be able to evaluate these choices. This time should be at least the time limit; however, to penalise these bad choices, in our implementation, we set the time to twice the time limit in such cases.

\section{Improvements when using the Classification Paradigm}
\label{sec:Classification}

In this section, we work with the supervised ML classification paradigm, most commonly used in the literature.   
First. in \Cref{sec:creating classification intance}, the usual approach to the creation of inputs for the ML model is described.  In \Cref{sec:exploring classification dataset} some problems in the created classification dataset is identified, and in \Cref{sec:improving datasets} some tools are introduced to solve these problems. In \Cref{sec:replicating methodology} the misleading results that can be obtained if these issues are not detected are presented, and in the following subsections a series of improvements to the classification model are presented to overcome these difficulties.  
The first improvement solves the imbalanced problem in the training dataset, the second increases the size of the training dataset, and the last reduces the number of features describing an instance while preserving most of the information these features carry.  These contributions were largely presented in the SC-Square 2023 workshop paper \cite{delrioDataAugmentationMathematical2023}.

\subsection{Creating a classification instance}\label{sec:creating classification intance}

A classification instance in supervised learning consists of a pair: an input and its label. In our case, from each set of polynomials in three variables, we must find its optimal variable ordering label.  We also need to create an embedding that can be taken as input by a ML model.

\subsubsection{Creating the inputs}
\label{subsec:classification_input}

The task here is to represent a set of polynomials in three variables as a list of features. In \Cref{sec:creating embeddings} we described a methodology for creating a list of features, which we will denote by $f_x(S)$ for a given variable $x$ and a given set of polynomials $S$.

To represent a three-variable set of polynomials $S\in\RR[x_1,x_2,x_3]$ to the classification model, we will therefore use the lists of features created with each variable appended to each other, placing the features related to the variable $x_1$ first, then those related to $x_2$ and finally those related to $x_3$. 

\subsubsection{Creating the labels}

The labels correspond to the possible variable orderings that a CAD can use for a set of polynomials $S\in\RR[x_1,x_2,x_3]$. We will use as a label the variable ordering for which building a CAD took the shortest time. We will use the notation $\succ_{ijk}$ to refer to ordering $x_i\succ x_j\succ x_k$.

\subsubsection{Illustrative example}
\label{subsec:illustrative_example}

\begin{figure}[ht]
    \centering
    \includegraphics[width=\textwidth]{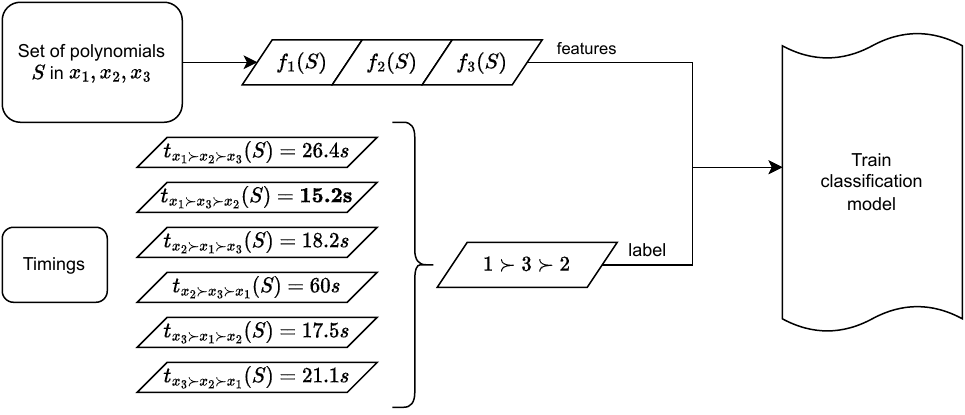}
    \caption{Classification training workflow for the example of \Cref{subsec:illustrative_example}}
    \label{fig:train_classification}
\end{figure}

For illustration purposes, we will consider a hypothetical example that produces the timings to build a three-variable CAD indicated in Table \ref{tab:orderings}.  Figure \ref{fig:train_classification} then illustrates how we create ML classification instances from our working example.  This will serve as a comparison for the alternative methodologies presented later.

\begin{table}[ht]
\def\arraystretch{1.5}
\setlength\tabcolsep{0.2cm}
\centering
\caption{Hypothetical timings to build a CAD for a set of polynomials with six possible variable orderings}
\label{tab:orderings}
\begin{tabular}{|c|c|}
\hline
\multicolumn{1}{|l|}{\textbf{Ordering}} & \textbf{Timing } \\ \hline
$\succ_{123}$  & 22.16s    \\ \hline
$\succ_{132}$  & 17.14s    \\ \hline
$\succ_{213}$  & Timeout 30s    \\ \hline
$\succ_{231}$  & 24.87    \\ \hline
$\succ_{312}$  & \textbf{16.06}    \\ \hline
$\succ_{321}$  & 22.58    \\ \hline
\end{tabular}
\end{table}

\subsubsection{Reflection on the number of variables}

Note that the length of the feature array describing the set of polynomials
and the number of labels (possible variable orderings) depends on the number of variables. Therefore, a model trained on three variable sets 
will not be useful for sets of polynomials with a higher number of variables. In \Cref{sec:RegressionOrderings} we will eliminate the dependence the output has on the number of variables, and in \Cref{sec:regression variables} we will further eliminate the dependence of the feature array with respect to the number of variables, creating a strategy that can be used to choose a variable ordering to build a CAD for sets of polynomials of an arbitrary number of variables.

\subsection{Imbalance in the dataset}\label{sec:exploring classification dataset}

As observed recently in \cite{delrioDataAugmentationMathematical2023} and \cite{hesterAugmentedMetiTarskiDataset2023} an analysis on the classification dataset described in \Cref{sec:creating classification intance} reveals that the dataset obtained is imbalanced: there are far more pairs of labels with ordering $\succ_{123}$ than any of the others, as can be observed in \Cref{fig:instances imbalanced}.  
Such an imbalance is a problem for ML, because models trained on an imbalanced dataset will be biased towards the majority class. Moreover, this problem may go unnoticed if the testing dataset is also imbalanced in a similar way, reporting misleading results that will not generalise when testing on a balanced dataset. 

There is no reason to expect real-world data to have such an imbalance and so it is essential to balance the data prior to ML to avoid model bias.    
There are various approaches: e.g., we could oversample the minority classes (giving instances in the minority classes multiple times to the model), or undersample the majority classes (not giving all instances in the majority classes to the model).  Instead, we use a data augmentation technique to generate more instances of the minority classes.

\subsection{Improving the datasets with data augmentation}\label{sec:improving datasets}

As introduced in \Cref{subsec:intro_data_augmentation}, data augmentation consists of generating new data instances from existing ones and is a common technique in computer vision.

As a motivating example, let us imagine that we have a dataset of arrows: 56 arrows pointing up, 35 left, 4 downward, and 175 right.  Like our CAD dataset, this dataset is very imbalanced; any model trained on it would likely have a bias towards predicting that the arrow points to the right and against predicting that the arrow points downward. 

It is clear to any human that a picture of an arrow pointing to the right that is rotated 90 degrees clockwise results gives a picture of an arrow pointing downward. Similar to this we could rotate by $180^{\circ}$ or $270^{\circ}$ clockwise to give arrows facing left and up, respectively.  
Editing images in this way can be very useful because it allows us to obtain a balanced dataset from the imbalanced dataset: simply randomly rotating each image by $0^{\circ}, 90^{\circ}, 180^{\circ}$ or $270^{\circ}$.

Our instances consist of sets of polynomials, for example, $\{x_1^2-x_2,x_3^3-1\}$ for which we can determine, by computing and comparing CADs, that the optimal variable ordering to compute a CAD for this set is $x_2\succ x_1\succ x_3$. Now observe that simply by swapping the names of the variables $x_1$ and $x_2$ we may obtain the new set of polynomials $\{x_2^2-x_1,x_3^3-1\}$, in which we know, without any further CAD computation, that the optimal variable ordering is $x_1\succ x_2\succ x_3$.

\begin{figure}[ht]
    \centering
    \includegraphics[width=\textwidth]{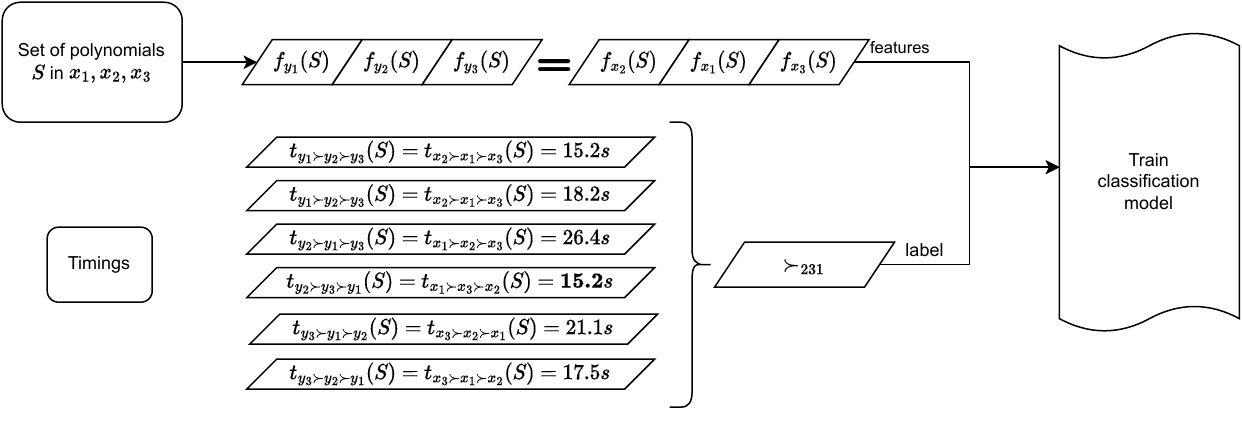}
    \caption{Classification training workflow with data augmentation: $S'$ has been created by renaming $x_1$ to $y_2$, $x_2$ to $y_1$ and $x_3$ to $y_3$ in $S$.}
    \label{fig:augmentation}
\end{figure}

This process of how an instance with a different label can be obtained from an existing instance is illustrated in \Cref{fig:augmentation} for our example of \Cref{subsec:illustrative_example}.

\subsubsection{Three datasets}

We will refer to the dataset described in \Cref{sec:dataset} as the ``\emph{imbalanced dataset}''.  
Similar to how the arrow image dataset can be balanced by randomly rotating, our dataset can be balanced by randomly permuting the variable names: we refer to the dataset obtained this was as the ``\emph{balanced dataset}''.  We then go further and fully augment the dataset by including all possible variable permutations, obtaining six instances from each original one: creating the ``\emph{augmented dataset}''. The distribution of instances to optimal orderings is illustrated in \Cref{fig:instances imbalanced} $-$ \Cref{fig:instances augmented}.

Note that in these two new datasets, instances are first separated in folds and then either balanced or augmented. This ensures that even when augmenting the dataset there is no data leakage.

\begin{figure}
\centering
\begin{tikzpicture}[scale=.85]
      \begin{axis}
      [  
          ybar,
          ylabel={\# Imbalanced training instances}, 
          xlabel={\ Ordering},  
          symbolic x coords={$\succ_{123}$, $\succ_{132}$, $\succ_{213}$, $\succ_{231}$, $\succ_{312}$, $\succ_{321}$}, 
          nodes near coords, 
          ymajorticks=false,
          ymin = 0,
          ymax = 1200,
          ]
          {
              \addplot coordinates 
              {
              ($\succ_{123}$,406)($\succ_{132}$,93)($\succ_{213}$,135)($\succ_{231}$,51)($\succ_{312}$,202)($\succ_{321}$,132)
              };
          }
      \end{axis}
  \end{tikzpicture} 
\caption{Instances in  
 the classes in the imbalanced classification dataset. \label{fig:instances imbalanced}}

\begin{tikzpicture}[scale=0.85]
      \begin{axis}
      [  
          ybar,
          ylabel={\# Balanced training instances}, 
          xlabel={\ Ordering},  
          symbolic x coords={$\succ_{123}$, $\succ_{132}$, $\succ_{213}$, $\succ_{231}$, $\succ_{312}$, $\succ_{321}$}, 
          nodes near coords, 
          ymajorticks=false,
          ymin = 0,
          ymax = 1200,
          ]
          {
              \addplot coordinates 
              {
              ($\succ_{123}$,188)($\succ_{132}$,170)($\succ_{213}$,153)($\succ_{231}$,158)($\succ_{312}$,188)($\succ_{321}$,162)
              };
          }
      \end{axis}
  \end{tikzpicture} 
\caption{Instances in  
 the classes in the balanced classification dataset. \label{fig:instances balanced}}

\begin{tikzpicture}[scale=0.85]
      \begin{axis}
      [  
          ybar,
          ylabel={\# Augmented training instances}, 
          xlabel={\ Ordering},  
          symbolic x coords={$\succ_{123}$, $\succ_{132}$, $\succ_{213}$, $\succ_{231}$, $\succ_{312}$, $\succ_{321}$}, 
          nodes near coords, 
          ymajorticks=false,
          ymin = 0,
          ymax = 1200,
          ]
          {
              \addplot coordinates 
              {
              ($\succ_{123}$,1019)($\succ_{132}$,1019)($\succ_{213}$,1019)($\succ_{231}$,1019)($\succ_{312}$,1019)($\succ_{321}$,1019)
              };
          }
      \end{axis}
  \end{tikzpicture} 
\caption{Instances in 
the classes in the augmented classification dataset. \label{fig:instances augmented}}
\end{figure}

\subsubsection{Other ideas to augment the dataset}

In computer vision, not only rotations are used as a tool to balance and augment datasets, but also other operations such as flipping or zooming.

The equivalent of flipping an image for a set of polynomials would be making changes of variables of the form $y_i=(-1)^{j_i}x_i$, where $j_i$ is an integer for all $i$. This change of variables would not change the computation of the CAD and therefore the label would remain the same; however, it does not affect the features used to describe the set of polynomials; so the instances obtained by this change of variables would be just copies of existing instances.

The equivalent of zooming an image for a set of polynomials could be making changes of variables of the form $y_i=j_ix_i+k_i$, where $j_i,k_i\in\RR$ for all $i$. These changes of variables do affect the features used to describe the set of polynomials, mainly because they tend to change the sparsity of the variables, one of the features used, so they would definitely create new instances. However, these changes in sparsity also affect the computations of resultants and discriminants needed to build a CAD, so the label of the instance may change; meaning that constructing CADs for all possible orderings would be necessary to find the real label of the instance.

\subsection{Replicating prior work}\label{sec:replicating methodology}

\begin{figure}[ht]
    \centering
    \includegraphics[width=0.7\textwidth]{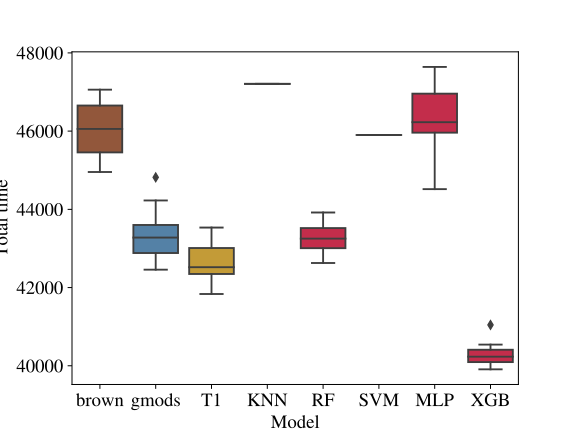}
    \caption{Existing heuristics and ML models trained on the biased dataset, evaluated over the biased dataset. See \Cref{sec:heuristics} for the definitions of \texttt{gmods}, \texttt{brown} and \texttt{T1}, and see \Cref{sec:training models} for definitions of the other acronyms.}
    \label{fig:florescu_vs_heuristics}
\end{figure}

To replicate the work done by Florescu and England \cite{florescuImprovedCrossValidationClassifiers2020} we used the biased training dataset to train ML classifiers, and we tested it in the biased testing dataset.  As can be seen in \Cref{fig:florescu_vs_heuristics}, these models perform similarly to existing heuristics, some outperforming them.

However, these models are trained in a biased dataset, and so likely to learn to replicate these biases. The biased dataset contains a large percentage of instances with label $\succ_{123}$, and therefore, the models may simply learn to predict this class more often.  
So when these models are tested on a dataset that has the same bias, their performance may be misleading. 

\begin{figure}[ht]
    \centering
    \includegraphics[width=0.7\textwidth]{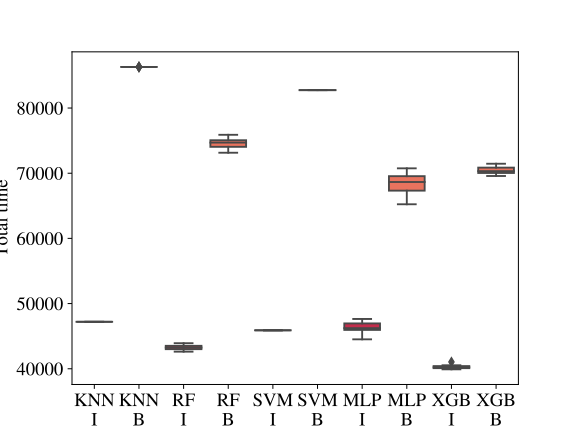}
    \caption{ML models trained on a biased dataset tested on both imbalanced (I) and balanced (B) datasets}
    \label{fig:test_in_bias_vs_balanced}
\end{figure}

To identify if this is the case, we may evaluate those models trained using the biased training dataset on the balanced testing dataset.  As expected, all models suffer a very significant performance drop, as illustrated in \Cref{fig:test_in_bias_vs_balanced}.

\subsection{ML classifiers obtained when balancing and augmenting the training dataset}
\label{sec:improving quality training}

So models trained imbalanced dataset perform poorly on a balanced one.  We will test whether using the balanced dataset for training the models will improve the performance: this should be the case since the models will no longer learn the biases they got when trained in an imbalanced dataset.  
We will also evaluate if using the augmented dataset described in \Cref{sec:improving datasets} offers further benefits:  it has six times the number of instances of the imbalanced and balanced datasets.  

The results in Tables \ref{tab:quality dataset knn} $-$ \ref{tab:quality dataset xgb} show the performance of models trained on our three datasets when evaluated on the balanced dataset and confirm our hypotheses.  Balancing the data avoided the biases and the additional data from further augmentation had a further positive effect.

\begin{table}[ht]
\centering

\begin{tabular}{|c|c|c|c|c|c|}
\hline
\textbf{Training dataset} & \textbf{Total time} & \textbf{Completed} & \textbf{Accuracy} & \textbf{Markup}\\
\hline
\csvreader[late after line=\\\hline]%
  {Tables/comparing_training_qualities_KNN-Classifier.csv}
  {TrainingQuality=\TrainingQuality,TotalTime=\TotalTime, Completed=\Completed, Accuracy=\Accuracy, Markup=\Markup}%
  {\TrainingQuality & \TotalTime & \Completed  & \Accuracy & \Markup}%
\hline
\end{tabular}
\caption{Metrics over the balanced testing dataset for the KNN classifier \label{tab:quality dataset knn}}

\vspace*{0.1in}

\begin{tabular}{|c|c|c|c|c|c|}
\hline
\textbf{Training dataset} & \textbf{Total time} & \textbf{Completed} & \textbf{Accuracy} & \textbf{Markup}\\
\hline
\csvreader[late after line=\\\hline]%
  {Tables/comparing_training_qualities_MLP-Classifier.csv}
  {TrainingQuality=\TrainingQuality,TotalTime=\TotalTime, Completed=\Completed, Accuracy=\Accuracy, Markup=\Markup}%
  {\TrainingQuality & \TotalTime & \Completed & \Accuracy & \Markup}%
\hline
\end{tabular}
\caption{Metrics over the balanced testing dataset for the MLP classifier. \label{tab:quality dataset mlp}}

\vspace*{0.1in}

\begin{tabular}{|c|c|c|c|c|c|}
\hline
\textbf{Training dataset} & \textbf{Total time} & \textbf{Completed} & \textbf{Accuracy} & \textbf{Markup}\\
\hline
\csvreader[late after line=\\\hline]%
  {Tables/comparing_training_qualities_RF-Classifier.csv}
  {TrainingQuality=\TrainingQuality,TotalTime=\TotalTime, Completed=\Completed,  Accuracy=\Accuracy, Markup=\Markup}%
  {\TrainingQuality & \TotalTime & \Completed & \Accuracy & \Markup}%
\hline
\end{tabular}
\caption{Metrics over the balanced testing dataset for the RF classifier. \label{tab:quality dataset rf} }

\vspace*{0.1in}

\begin{tabular}{|c|c|c|c|c|c|}
\hline
\textbf{Training dataset} & \textbf{Total time} & \textbf{Completed} & \textbf{Accuracy} & \textbf{Markup}\\
\hline
\csvreader[late after line=\\\hline]%
  {Tables/comparing_training_qualities_SVM-Classifier.csv}
  {TrainingQuality=\TrainingQuality,TotalTime=\TotalTime, Completed=\Completed, Accuracy=\Accuracy, Markup=\Markup}%
  {\TrainingQuality & \TotalTime & \Completed  & \Accuracy & \Markup}%
\hline
\end{tabular}
\caption{Metrics over the balanced testing dataset for the SVM classifier. \label{tab:quality dataset svm}}

\vspace*{0.1in}

\begin{tabular}{|c|c|c|c|c|c|}
\hline
\textbf{Training dataset} & \textbf{Total time} & \textbf{Completed} & \textbf{Accuracy} & \textbf{Markup}\\
\hline
\csvreader[late after line=\\\hline]%
  {Tables/comparing_training_qualities_XGB-Classifier.csv}
  {TrainingQuality=\TrainingQuality,TotalTime=\TotalTime, Completed=\Completed, Accuracy=\Accuracy, Markup=\Markup}%
  {\TrainingQuality & \TotalTime & \Completed  & \Accuracy & \Markup}%
\hline
\end{tabular}
\caption{Metrics over the balanced testing dataset for the XGB classifier. \label{tab:quality dataset xgb}}
\end{table}

\section{Changing from the Classification to the Regression Paradigm}
\label{sec:RegressionOrderings}

As we described in \Cref{subsec:intro_paradigms}, although prior work on ML to make choices for our problem has employed the classification paradigm, since the choices seek to minimise a continuous metric the problem may also be conceived as ML regression. We explore the effects of this paradigm change in this section: first describing how regression instances are created in \Cref{sec:creating regression intance},  then presenting an experimental comparison in \Cref{subsec:CvsR}.

\subsection{Creating a regression instance}\label{sec:creating regression intance}

\begin{figure}
    \centering
    \includegraphics[width=\textwidth]{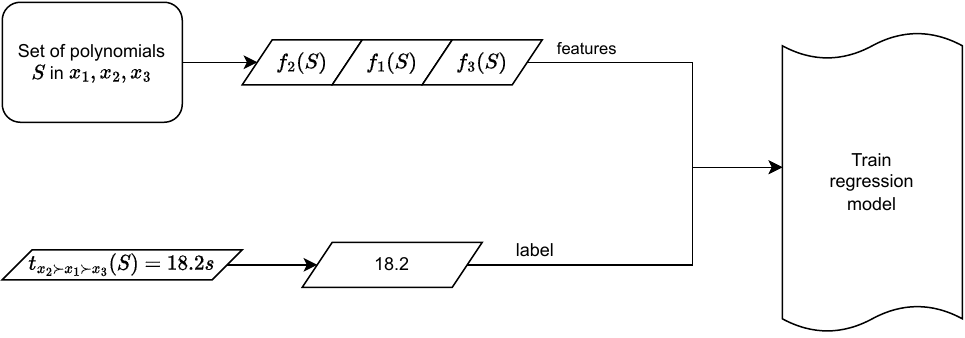}
    \caption{Regression training workflow showing how a single regression instance is created from just 1 of the 6 orderings for the example of \Cref{subsec:illustrative_example}; 5 further instances may be created similarly.}
    \label{fig:train_regression}
\end{figure}

A regression instance consists of a pair: an input and a target. In our case, the input is formed from the set of polynomials and variable ordering, and the target will be the time taken to build a CAD for these.  The embedding that we will use to encode the input will be formed by the same three vectors of features introduced previously in \Cref{subsec:classification_input} for classification.  However here, as illustrated in \Cref{fig:train_regression}, the order in which we append the vectors will be determined by the input variable ordering.  

This embedding eliminates all existing symmetries in the classification embedding, so it is not possible to augment this dataset as we did in \Cref{sec:improving datasets} for classification. However, the number of instances in the regression dataset is already as large as that of the augmented classification dataset since each classification instance becomes six regression instances.  Furthermore, the regression dataset encodes more information than the augmented classification dataset. Although instances in both datasets share the same features, their labels differ: the augmented classification dataset labels are binary, indicating only which ordering is the fastest, while the regression dataset labels contain numerical information about the exact time each ordering took.

\begin{figure}
    \centering
    \includegraphics[width=\textwidth]{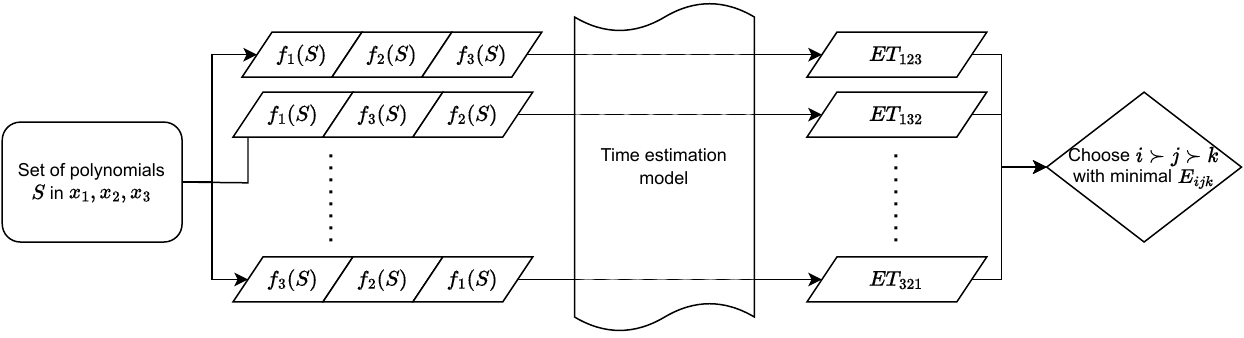}
    \caption{Overview of how the regression model is used.}
    \label{fig:test_regression}
\end{figure}

Our ML regression models will predict how long it takes to build a CAD for a given set of polynomials using a given variable ordering. However, this is not our final goal: we are interested in choosing variable orderings to build a CAD for a given set of polynomials as fast as possible. Thus, for a given set of polynomials in the testing dataset, we must ask the regression model to estimate how long it would take to build a CAD for each of the variable orderings, and then we will choose the ordering that was estimated to take the shortest time, as illustrated in \Cref{fig:test_regression}.  This final choice can then be evaluated using the same metrics introduced in \Cref{sec:testing models}.

\subsection{Results comparing classification with regression}
\label{subsec:CvsR}

\begin{figure}
    \centering
    \includegraphics[width=0.7\textwidth]{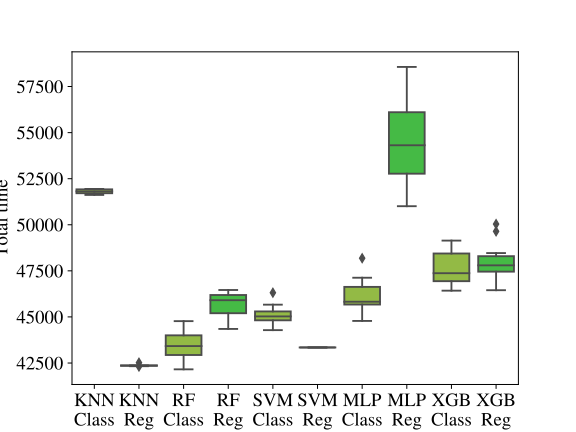}
    \caption{Comparing the total times to build CADs for the balanced dataset with the variable ordering suggested by classifiers (Class) and regressors (Reg)}
    \label{fig:classification_vs_regression}
\end{figure}

Figure \ref{fig:classification_vs_regression} compares the total time required to build the CADs in our test dataset using the variable ordering suggested by each model.  We see that the KNN, and SVM, models all perform better under the regression paradigm, while the RF, MLP, and XGBoost models perform worse.  Overall, the best performance on the dataset is achieved by a model under the regression paradigm.  So the results partially validate our hypothesis that regression can use extra information to improve performance on the choice, but with the substantial caveat that it is model dependent.
 
We also reflect that the scope of choices that can be made using regression is broader than those that can be made using classification. For example, when using CAD for quantifier elimination, we must project variables in the order of quantification but still have freedom to change order within quantification blocks (and the free variables).  In this case, if a classifier returns an option that does not satisfy these restrictions the recommendation is useless; however, using the regression paradigm, one can simply take the best choice that meets the restrictions.

\subsection{Limitations with respect to number of variables}
\label{sec:not more than three regression}

The number of possible outputs the classification models have to choose from is factorial of the number of variables.  A classification model trained for $n$ variables cannot study a problem with more variables; and although the methodology could generalise (i.e., we may train new classifiers with a dataset of problems in more variables), the factorial growth means this will get harder quickly.  

In comparison the regression models presented in this section always produced a single real number regardless of the number of variables.  The difficulty in generalising to more variables will be the embedding used to input sets of polynomials into both the regression and classification models: this will still grow with the number of variables, but only linearly (27 times the number of variables).

If we want to choose variable orderings for an arbitrary number of variables, then ideally we should find an embedding that has an invariant size with respect to the number of variables. We consider one approach to this in \Cref{sec:regression variables}, in which our model takes always a fixed-size input of only 27 features.

\section{Using Regression to Rank Variables}
\label{sec:regression variables}

In \Cref{sec:RegressionOrderings} we trained regression models to estimate how long it would take to build a CAD for a problem with a given variable ordering.  In this section we will use ML to rank the variables themselves instead of the variable orderings, allowing for an easier generalisation to more variables.  

This is similar to how the human-designed heuristics in \Cref{sec:heuristics} choose a variable ordering. Human heuristics assign some measure of complexity to the variables and pick the simpler variables first in the ordering, but their measure of complexity is defined by the creator of the heuristic and, therefore, is fixed. Instead, we will train ML models so that they learn their own measures of complexity, and we will pick the variables that the model considers to be simpler first in the ordering.
This shares a similarity with Chen et al. (2020) \cite{chenVariableOrderingSelection2020} who also proposed to train models to choose variables instead of orderings.  However, they still trained classification models and thus suffer from the limitations discussed above, being limited in their approach to sets of polynomials with less than ten variables. 

Moreover, instead of choosing the whole variable ordering using only the original polynomials, once the first variable has been chosen, we will use the projected set to choose the next variable. This allows us to make our choices using the most recent information.

\subsection{Creating instances for regression on variables}\label{sec:approaches regression variable instances}

Our instances now concern the polynomials and a single variable.  For an embedding we use just the features for that variable, i.e. $f_{i}(S)$: the features describing variable $x_i$ in a set of polynomials $S$.

\begin{figure}
     \centering
     \includegraphics[width=\textwidth]{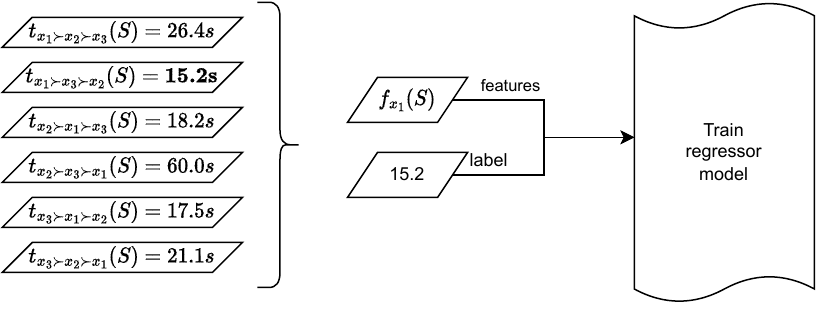}
     \caption{Illustration of how the instance for variable $x_1$ is encoded for the example of \Cref{subsec:illustrative_example}.  Note that two further instances for the other variables will be created similarly. }
     \label{fig:train regression variable 1}
\end{figure}

In \Cref{sec:RegressionOrderings} we trained ML models to learn the complexity of a variable ordering and there was a clear output to use: the time necessary to build a CAD using such variable ordering.  
In this section, for estimating the complexity of a variable, we have available the timings of the orderings that start with that variable: in our three-variable case study there are two possible orderings that begin with each variable. We use the quickest of these to label the instance, as illustrated in \Cref{fig:train regression variable 1}.

\subsection{Results}

Tables \ref{tab:estrategies knn} $-$ \ref{tab:estrategies xgb} compare the classification paradigm (trained with the augmented dataset), the regression paradigm on the orderings, and the regression paradigm on the variables.  Interestingly, we see that the two models which did poorest under the regression paradigm to pick an ordering MLP and XGBoost do better under this alternative regression paradigm (while the others do worse).  The lowest timing overall remains the KNN under the previous regression paradigm, but given the wider applicability of this alternative paradigm we think it worthy of consideration.

\begin{figure}[ht]
  \centering
\begin{subfigure}{.48\textwidth}
  \centering
  \includegraphics[width=0.9\textwidth]{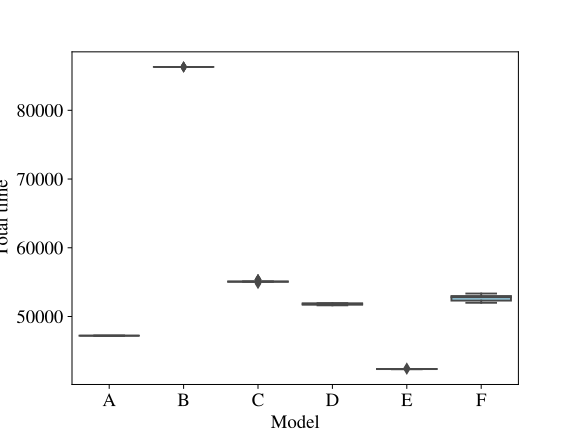}
    \caption{KNN models.}
  \label{fig:all strategies KNN}
\end{subfigure}%
\begin{subfigure}{.48\textwidth}
  \centering
  \includegraphics[width=0.9\textwidth]{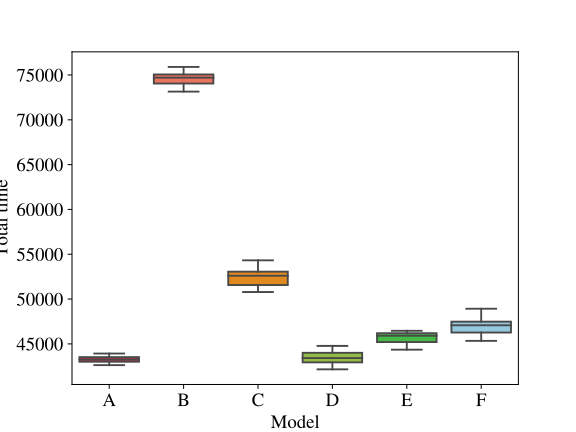}
    \caption{RF models.}
  \label{fig:all strategies RF}
\end{subfigure}
\begin{subfigure}{.48\textwidth}
  \centering
  \includegraphics[width=0.9\textwidth]{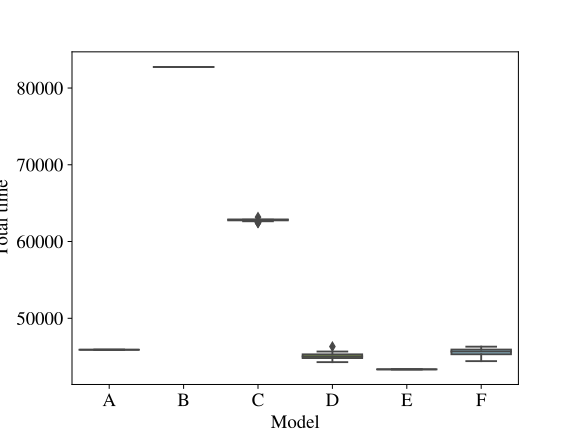}
    \caption{SVM models.}
  \label{fig:all strategies SVM}
\end{subfigure}
\begin{subfigure}{.48\textwidth}
  \centering
  \includegraphics[width=0.9\textwidth]{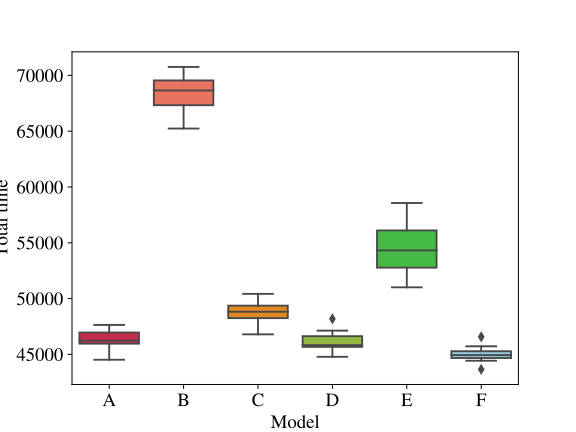}
    \caption{MLP models.}
  \label{fig:all strategies MLP}
\end{subfigure}
\begin{subfigure}{.48\textwidth}
  \centering
  \includegraphics[width=0.9\textwidth]{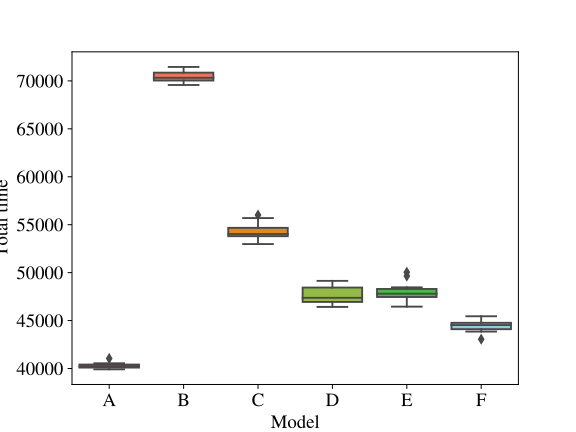}
    \caption{XGBoost models.}
  \label{fig:all strategies XGB}
\end{subfigure}
\caption{The total times achieved by models on the balanced testing dataset under different strategies \label{fig:all strategies}}
\end{figure}

In Figure \ref{fig:all strategies} we show the spread of timings for all the key strategies presented in this paper, represented by letters as below.  They show that for all models the original approach of training on an unbalanced dataset does not generalise to a balanced dataset (A vs B); but that this performance is regained and sometimes exceeded by the new approaches of this paper (C $-$ F).  The optimal paradigm does seem to depend on the ML model type being used.

\vspace*{0.1in}

\begin{tabular}{|c|c|c|c|}
\hline
\textbf{Short name} & \textbf{Training dataset} & \textbf{Testing dataset} & \textbf{Paradigm}\\\hline
A & Imbalanced (see \Cref{sec:exploring classification dataset}) & Imbalanced & Classification\\\hline
B & Imbalanced & Balanced & Classification\\\hline
C & Balanced (see \Cref{sec:improving datasets}) & Balanced & Classification\\\hline
D & Augmented (see \Cref{sec:improving datasets}) & Balanced & Classification\\\hline
E & Regression orderings (See \Cref{sec:creating regression intance}) & Balanced & Regression\\\hline
F & Regression variables (see \Cref{sec:regression variables})  & Balanced & Regression\\\hline
\end{tabular}

\begin{table}[!ht]
\centering

\begin{tabular}{|c|c|c|c|c|c|}
\hline
\textbf{Strategy} & \textbf{Total time} & \textbf{Completed} & \textbf{Accuracy} & \textbf{Markup}\\
\hline
\csvreader[late after line=\\\hline]%
  {Tables/comparing_ClassAugmented_vs_RegOrderings_vs_RegVariables-KNN.csv}
  {Strategy=\Strategy,TotalTime=\TotalTime, Completed=\Completed, Accuracy=\Accuracy, Markup=\Markup}%
  {\Strategy & \TotalTime & \Completed  & \Accuracy & \Markup}%
\hline
\end{tabular}
\caption{Metrics for strategies using a KNN model over the balanced dataset. \label{tab:estrategies knn}}

\vspace*{0.1in}

\begin{tabular}{|c|c|c|c|c|c|}
\hline
\textbf{Strategy} & \textbf{Total time} & \textbf{Completed} & \textbf{Accuracy} & \textbf{Markup}\\
\hline
\csvreader[late after line=\\\hline]%
  {Tables/comparing_ClassAugmented_vs_RegOrderings_vs_RegVariables-RF.csv}
  {Strategy=\Strategy,TotalTime=\TotalTime, Completed=\Completed, Accuracy=\Accuracy, Markup=\Markup}%
  {\Strategy & \TotalTime & \Completed  & \Accuracy & \Markup}%
\hline
\end{tabular}
\caption{Metrics for strategies using a RF model over the balanced dataset. \label{tab:estrategies rf}}

\vspace*{0.1in}

\begin{tabular}{|c|c|c|c|c|c|}
\hline
\textbf{Strategy} & \textbf{Total time} & \textbf{Completed} & \textbf{Accuracy} & \textbf{Markup}\\
\hline
\csvreader[late after line=\\\hline]%
  {Tables/comparing_ClassAugmented_vs_RegOrderings_vs_RegVariables-SVM.csv}
  {Strategy=\Strategy,TotalTime=\TotalTime, Completed=\Completed, Accuracy=\Accuracy, Markup=\Markup}%
  {\Strategy & \TotalTime & \Completed  & \Accuracy & \Markup}%
\hline
\end{tabular}
\caption{Metrics for strategies using an SVM model over the balanced dataset. \label{tab:estrategies svm}}

\vspace*{0.1in}

\begin{tabular}{|c|c|c|c|c|c|}
\hline
\textbf{Strategy} & \textbf{Total time} & \textbf{Completed} & \textbf{Accuracy} & \textbf{Markup}\\
\hline
\csvreader[late after line=\\\hline]%
  {Tables/comparing_ClassAugmented_vs_RegOrderings_vs_RegVariables-MLP.csv}
  {Strategy=\Strategy,TotalTime=\TotalTime, Completed=\Completed, Accuracy=\Accuracy, Markup=\Markup}%
  {\Strategy & \TotalTime & \Completed  & \Accuracy & \Markup}%
\hline
\end{tabular}
\caption{Metrics for strategies using a MLP model over the balanced dataset. \label{tab:estrategies mlp}}

\vspace*{0.1in}

\begin{tabular}{|c|c|c|c|c|c|}
\hline
\textbf{Strategy} & \textbf{Total time} & \textbf{Completed} & \textbf{Accuracy} & \textbf{Markup}\\
\hline
\csvreader[late after line=\\\hline]%
  {Tables/comparing_ClassAugmented_vs_RegOrderings_vs_RegVariables-XGB.csv}
  {Strategy=\Strategy,TotalTime=\TotalTime, Completed=\Completed, Accuracy=\Accuracy, Markup=\Markup}%
  {\Strategy & \TotalTime & \Completed  & \Accuracy & \Markup}%
\hline
\end{tabular}
\caption{Metrics for strategies using an XGB model over the balanced dataset. \label{tab:estrategies xgb}}

\end{table}

\section{Final Thoughts}
\label{sec:conclusion}

\subsection{Conclusions on data balance and augmentation}\label{sec:conclusions}
\label{sec:conc}

\subsubsection{The importance of data exploration}

We found in Section \ref{sec:Classification} that this well known dataset was imbalanced for our problem and that if we trained classifiers with such imbalanced data they could not perform well on balanced data.
This demonstrates the importance of exploring the data before moving to ML.  Imbalance is not always inappropriate for ML: some applications will naturally have imbalanced data, and ML models should be aware of this.  However, in our case, we seek heuristics for choosing variable orderings for CAD applied in general, and there is little rationale to suppose that general CAD applications favour one ordering over another\footnote{except perhaps the existence of the SMT-LIB data!}.  Thus, our advice is to ensure that ML models for such applications are trained on balanced data. 

\subsubsection{Recovering performance through balancing and augmentation}

The good news is that ML performance can be improved simply by training on balanced data, reducing the total time by 28\% on average, re-validating the value of the data science-led approach.  
A key conclusion here is that it is beneficial to go further and use maximum data augmentation: all models benefited from this beyond just balancing the data, no matter which dataset they are tested on.  Training in an augmented dataset of six times the size of the original dataset reflected in an increase in accuracy of 40\% on average and a reduction of 38\% in the total time on average.  In fact, the performance lost from the original imbalanced case is almost fully recovered this way.  

\subsubsection{Comparison with the work of Hester et al. (2023)}
\label{sec:comparison Hester}

Some of these conclusions were also drawn in a recent paper \cite{hesterAugmentedMetiTarskiDataset2023}. Table 2 in \cite{hesterAugmentedMetiTarskiDataset2023} presents the accuracies of trained models on different datasets: note that both their ``\emph{Training Set 2}'' and ``\emph{Dataset 1}'' contain instances in which the models have been trained, meaning that the former is the most appropriate column for evaluation in that table: that column shows results similar to ours.

We note that the original dataset in \cite{hesterAugmentedMetiTarskiDataset2023} contained 6895 instances, while in our paper the initial dataset only contained 1019 instances. This is because, even though both datasets have the same ultimate source (the SMT-LIB), our dataset had been stripped of potential duplicate instances (sets of polynomials whose CAD tree structures are potentially identical for every variable ordering), as described in detail in \Cref{sec:dataset}.  We view this as a necessary step to make meaningful use of the QF\_NRA section of the SMT-LIB where there are many \emph{very} similar problems.

This comparison with \cite{hesterAugmentedMetiTarskiDataset2023} shows that the size alone of the dataset is not what matters (since \cite{hesterAugmentedMetiTarskiDataset2023} has a similar accuracy to the models presented here despite training with much more data).  Rather, it is the \emph{number of qualitatively different problems within the dataset}. That is, there is little benefit to including multiple very similar problems.  It may seem that data augmentation adds no new information, but since the ML models are not aware of these symmetries by exposing them with augmentation, we actually give them access to this information.

\subsection{Conclusions on ML paradigms}

Despite the fact that the regression dataset contains more information than the classification one, such as the specific timings of each ordering, this did not always lead to improvement in the performance of regression models compared to classification models: it seemed that some models were better able to do this than others and this choice of paradigm should be optimised along with model and hyperparameter selection in the ML workflow.  

The overall lowest timings were achieved by one of the new regression approaches.  The differences in performance were not huge:  as argued in \Cref{subsec:CvsR}, we suggest that more important than performance statistics is the wider scope of choices that can be made in symbolic computation using the new regression methodologies.  
Consider for example an application to $S$-pair selection in Buchberger's algorithm:  we do not know in advance of a decision how many pairs there will be to choose from, but this may be tackled using the paradigm presented in \Cref{sec:regression variables}, and still under a supervised learning paradigm rather than using reinforcement learning as in \cite{peiferLearningSelectionStrategies2020}.

\subsection{Future work}
\label{sec:future work}

A key area for future work could be the embeddings: those we use still ignore much information from the polynomials such as all coefficient information.  Jia et al. (2023) \cite{jiaSuggestingVariableOrder2023} already proposed a graph embedding that allows inputting a lot of information about a set of polynomials into a GNN model, and they experience a performance similar to that of the heuristic gmods in the SMT-LIB dataset and even beat this heuristic in a randomly generated dataset.

Given the success of this data augmentation, an obvious area for future work is to look for additional augmentation techniques.  Returning to the computer vision analogy: rotations are not the only augmentation tool; there are others also, e.g. mirror reflections. Regarding mathematical objects, a corresponding augmentation technique may be substituting a variable with its negative, which would create a new instance without the need for any further labelling. We could also consider more involved variable transformations; however, these would most likely require additional CAD computations for data labelling, which is the most expensive part of this whole process.

Finally, we note that all these ideas would generalise easily to variable ordering choice for the other decision procedures of non-linear real arithmetic commonly found in the wider toolchains of the SC$^2$ community. Further, the lessons we illustrated on the importance of exploratory data analysis and paradigm choice would be useful to many other choices available in symbolic computation.

\subsection*{Data Access Statement}

This work in this paper formed part of  Tereso del R\'{i}o's PhD thesis.  All code used for the experiments reported on in this paper (including to generate the results images ) can be found within the data release for that thesis here:  \url{https://zenodo.org/doi/10.5281/zenodo.10834972}.

\subsection*{Acknowledgments}

Tereso del Río is supported by Coventry University and a travel grant from the London Mathematical Society (LMS).  Matthew England is supported by UKRI EPSRC Grant EP/T015748/1, \emph{Pushing Back the Doubly-Exponential Wall of Cylindrical Algebraic Decomposition} (the DEWCAD Project).

\bibliographystyle{plainurl}

\end{document}